\documentclass[showpacs,amsmath,amssymb,amsthm,amsfonts,11pt]{revtex4}

\usepackage{natbib}
\usepackage{graphicx}
\DeclareGraphicsExtensions{.jpg,.jpeg,.pdf,.png,.mps,.eps,.ps,.gif}
\usepackage{color}
\newcommand{\bminil}[1]{\begin{minipage}[l]{#1 \textwidth}}
\newcommand{\bminir}[1]{\begin{minipage}[r]{#1 \textwidth}}
\newcommand{\bminic}[1]{\begin{minipage}[c]{#1 \textwidth}}
\newcommand{\emini}{\end{minipage}}
\newcommand{\EQL}{\begin{equation}\label}
\newcommand{\EQ}{\begin{equation}}
\newcommand{\EN}{\end{equation}}
\newcommand{\BFG}{\begin{figure}}
\newcommand{\EFG}{\end{figure}}
\newcommand{\ITM}{\begin{itemize}}
\newcommand{\ITN}{\end{itemize}}
\newcommand{\p}{\partial}

\newcommand{\bk}{\mbox{\boldmath$k$}}

\newcommand{\bS}{\mbox{\boldmath$S$}}

\newcommand{\bu}{\mathbf u}
\newcommand{\bv}{\mbox{\boldmath$v$}}

\newcommand{\bx}{\mbox{\boldmath$x$}}

\newcommand{\bomega}{\mbox{\boldmath$\omega$}} 
 
\newcommand{\bSigma}{\mbox{\boldmath$\Sigma$}} 

\newcommand{\refsoff}[1]{}

\newcommand{\etall}{{\it et al.}}
\newcommand{\biband}{\&~}
\newcommand{\authone}[2]{#2 #1,}
\newcommand{\authtwo}[4]{#2 #1, #4 #3, }
\newcommand{\auththr}[6]{#2 #1,#4 #3,#6 #5, }
\newcommand{\authfour}[8]{#2 #1,#4 #3,#6 #5,#8 #7, }
\newcommand{\authmanytwo}[4]{#2 #1,#4 #3, }
\newcommand{\authmanythr}[6]{#2 #1,#4 #3,#6 #5,}

\newcommand{\yjour}[6]{#6~{#2}~#1;{#3}:#4#5.}
\newcommand{\ybook}[3]{ {\em #2} #3; #1 }

\begin{document}
\title{Incompressible hydrodynamic turbulence from a chain reaction of
vortex reconnection events}
\author{Robert M. Kerr}
\affiliation{Department of Mathematics, 
University of Warwick, Coventry CV4 7AL, United Kingdom}

%\dochead{Extreme Events}
\begin{abstract}
From a new anti-parallel initial condition using long vortices, three-dimensional 
turbulence forms after two reconnection steps and the formation of at least one set of 
vortex rings for both quantum and incompressible vortices. The long domain allows
multiple reconnections, which enhance vortex stretching rates and the generation of 
small-scale vortex structures within the vortex rings. For the Navier-Stokes vortices, 
further new features are a profile less likely to shed vortex sheets and an improved 
mapping the direction of the vorticity onto the three-dimensional mesh. 
The vortices evolve via the following steps: First, until the first reconnection, 
dynamics largely consistent with how vortices attract with a possible singularity
in the Euler equations. Second, vortex reconnection where the symmetry planes meet. 
Third, a series of vortex rings, with the stretching at each following set of 
reconnections leading to the new reconnections and rings.
Half of the circulation 
reconnects into two ``bridges'', leaving two ``threads'', which the
extra stretching transforming the threads into spirals wrapped around the bridges.
It is argued that these spirals are the source of the observed $k^{-5/3}$ energy 
spectra and other statistics commonly associated with high Reynolds number 
turbulence.
\end{abstract}
\maketitle

\email{Robert.Kerr@warwick.ac.uk}

\begin{center} To appear in the {\it Procedia IUTAM} volume of papers for
Understanding Common Aspects of Extreme Events in Fluids, Dublin, 2-6 July 2012.
as {\it Fully developed hydrodynamic turbulence from a chain reaction of
reconnection events}.
%under \\ Fully developed hydrodynamic turbulence from a chain reaction of
%reconnection events 
\end{center}

\section{Introduction \label{sec:intro}}
\begin{figure}
\bminil{0.5}
\vspace{-10mm}
{\bf t=0, both cases.}\\
\includegraphics[scale=.12]{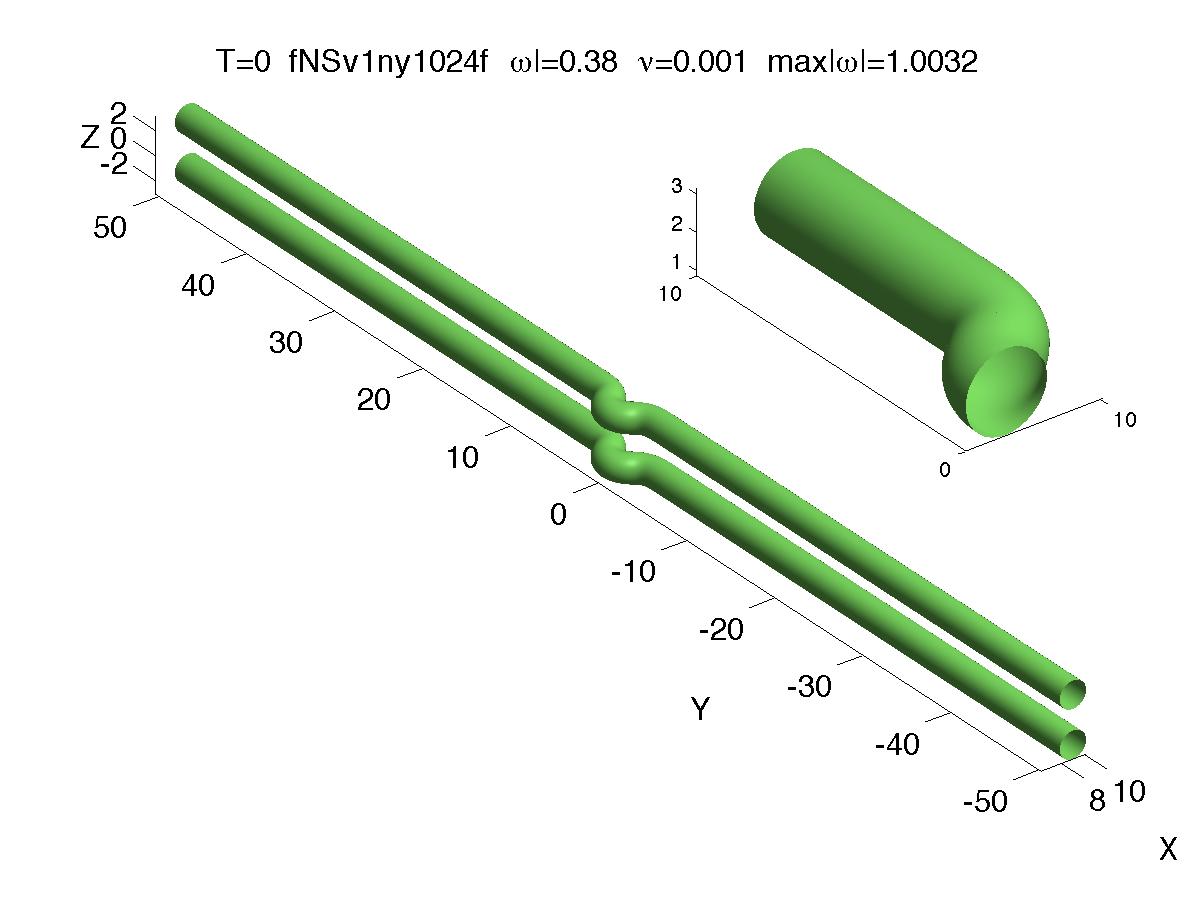}\\
{\bf GP transition: t=1.5.}\\
\includegraphics[scale=.40]{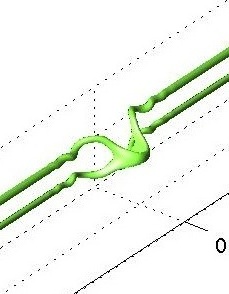}\\
{\bf GP rings: t=32.}\\
\includegraphics[scale=0.5]{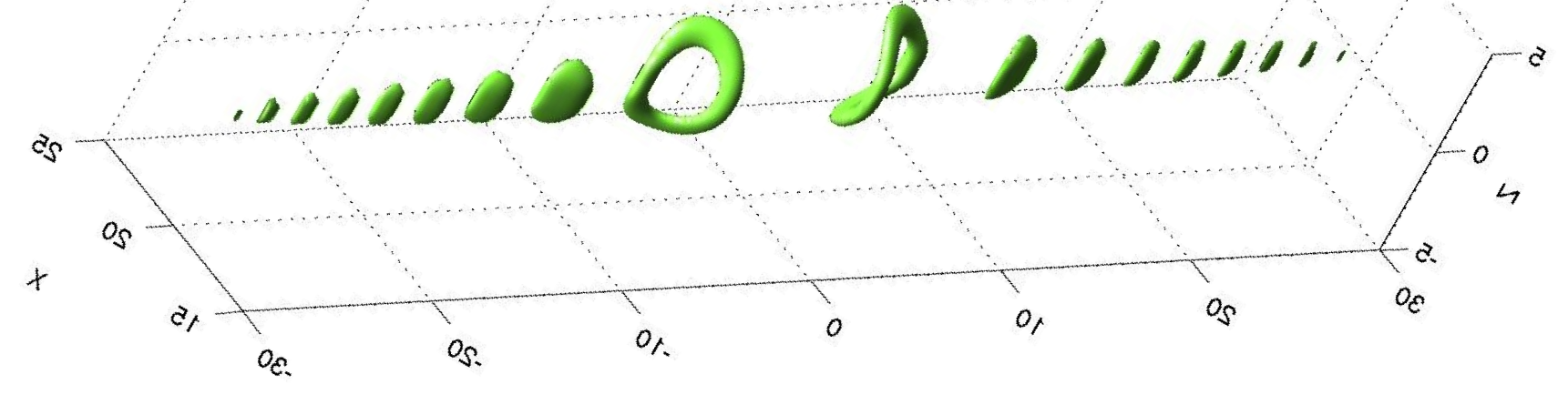}
\emini\bminir{0.5}
{\bf NS transition: t=16}\\
\includegraphics[scale=.12]{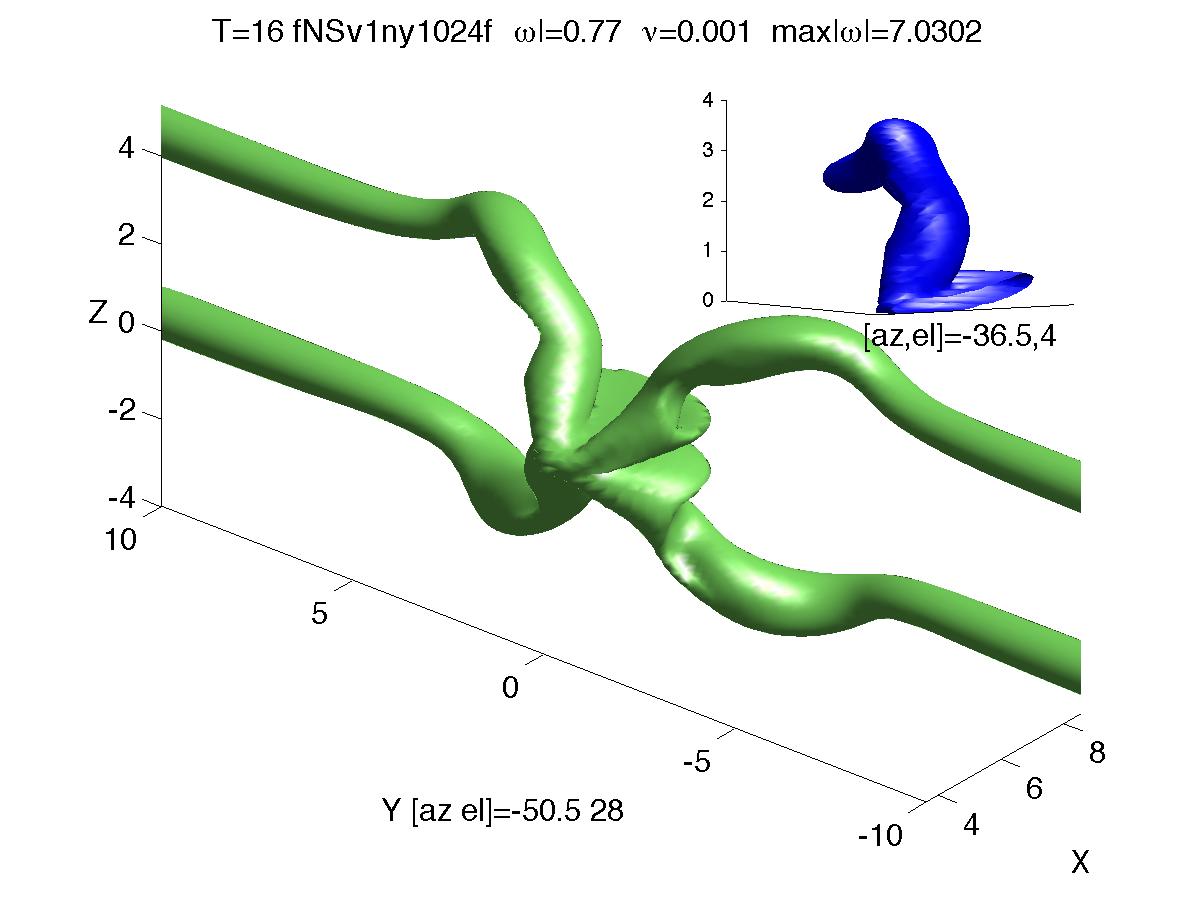} \\
{\bf NS first ring forms: t=96}\\
\includegraphics[scale=.15]{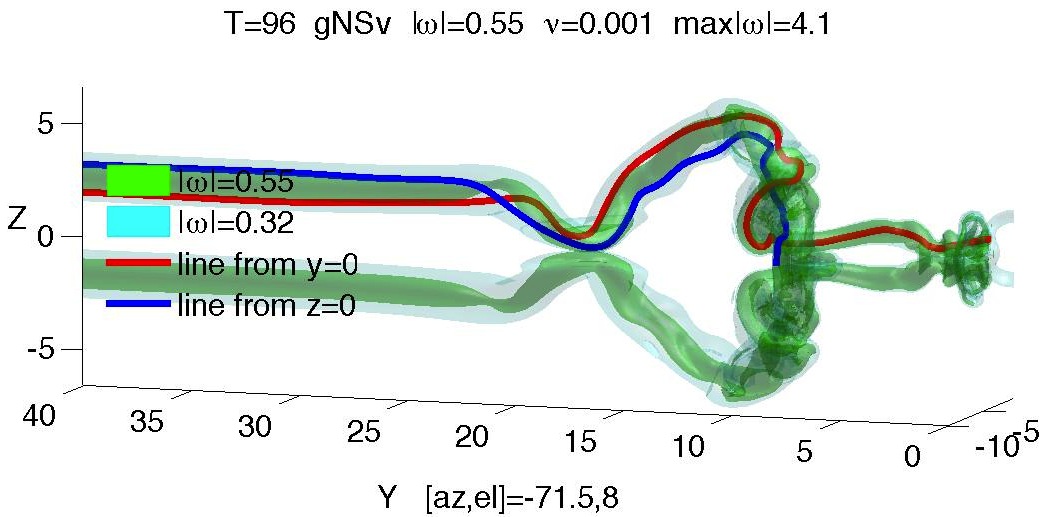} \\
{\bf NS turbulence: t=256}\\
\includegraphics[scale=.15]{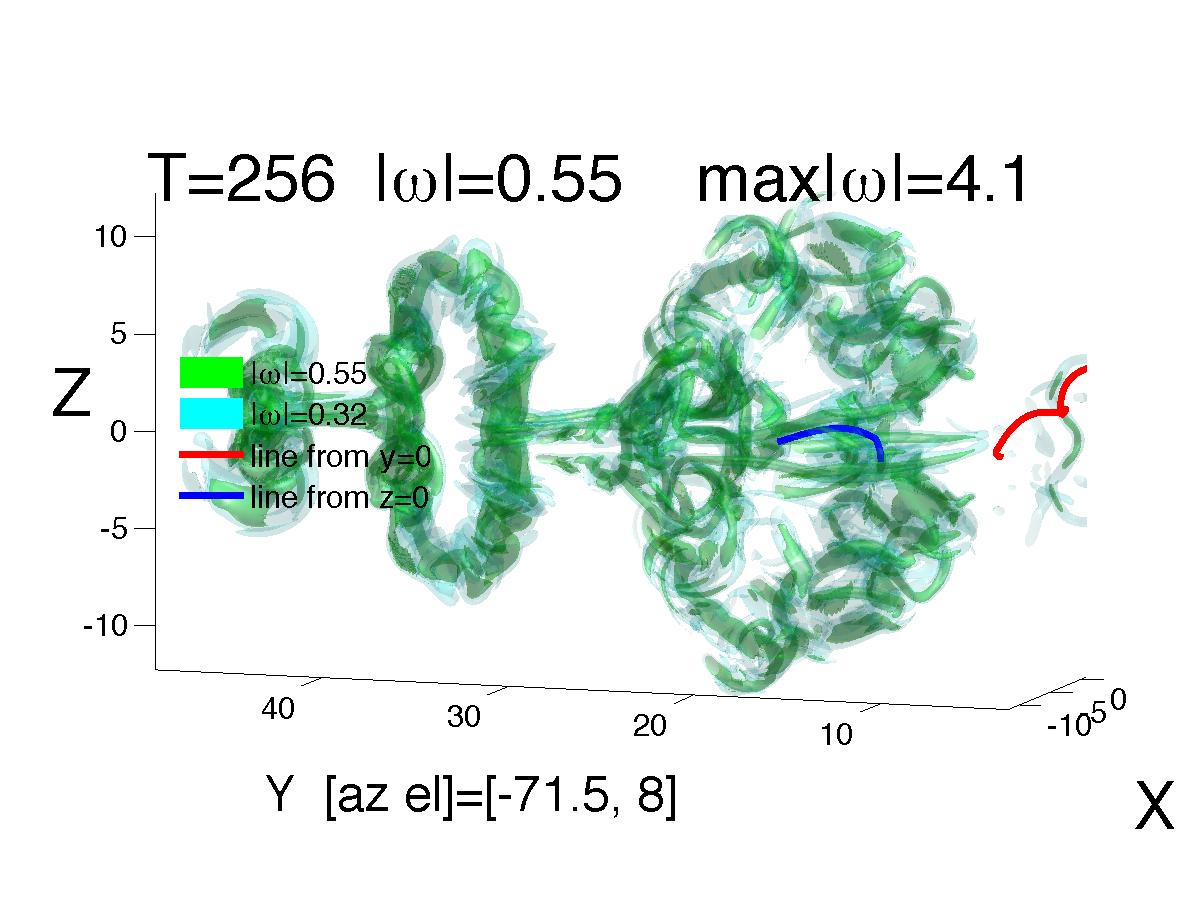}
\emini
\caption{Comparison of long anti-parallel evolution for Gross-Pitaevskii and
Navier-Stokes.  The objective was to have
an initial condition where vortex lines did not separate from the vicinity of the 
prescribed trajectory. In each case a series of reconnections results in the 
formation of a series of vortex rings.  The Navier-Stokes rings swirl due
to the incomplete reconnection of circulation. The steps are discussed in the
text.}
\label{fig:GPNSreconnect}
\end{figure}

This paper presents initial comparisons between new Navier-Stokes reconnection
calculations and recent quantum vortex reconnection calculations.  It is also
the first of a series of reports on hydrodynamic Navier-Stokes
and Euler simulations using a new set of anti-parallel vortex initial conditions.  
Figure \ref{fig:GPNSreconnect} shows the initial configuration of vorticity used in
all of these calculations, with vortices that are very long compared with 
the size of the initial perturbation.  For the classical calculations isosurfaces
of vortices and vortex lines are shown.  For the quantum case isosurfaces of
very low density are shown.

Preliminary analysis of these new simulations provides insight into these questions:
\ITM
\item[a)] Could a simple configuration of vortices generate turbulence promptly from 
a clearly identified series of steps? 
\item[b)] Can interacting vortices directly generate
a $k^{-5/3}$ energy spectrum without the usual statistical cascade arguments?
\item[c)] What is the role of temporal intermittency in determining
the higher-order statistics of turbulence? 
\ITN

The conclusion of this paper will be
that at least two reconnection steps are needed in the interaction between the
two vortices before the classical signatures of fully-developed Navier-Stokes 
turbulence appear. One of these signatures is a $k^{-5/3}$ energy spectrum.  The
development of -5/3 spectrum appears to be linked to the generation
of stretched spiral vortices \citep{Lundgren82}, which in turn grow out of
locally orthogonal vortices created by a series of vortex reconnection events,
a configuration that has been seen in some isotropic calculations 
\citep{Kerr85,HolmKerr07} just before they become fully turbulent.

This paper is organised as follows. First, the new initial condition. This will allow 
alternative numerical methods to reproduce the results given here. 
Third, the stages in the evolution of the vortex structures are described using 
figures \ref{fig:GPNSreconnect} and \ref{fig:T32}, starting from the first 
reconnection event and ending with a turbulent state of vortex rings with swirl. 
Some statistics for enstrophy growth will be provided to give a time-line for the
evolution.  After the final structure is described, spectra and then
vorticity moments will be discussed
to illustrate the evolution towards a turbulent state.

\section{Background: Classical versus quantum decaying turbulence}

Experimentally, a quantum vortex tangle, that is quantum turbulence, is
observed to decay \cite{Smithetal93,WalmsleyGolovetal07}
with the same power law as homogeneous, isotropic incompressible
turbulence in a periodic box and is observed to have a $k^{-5/3}$, as does
classical turbulence \cite{MaurerTabeling98,Rocheetal07}. Numerically, this can be
achieved using isotropic initial conditions \cite{Yepezetal09}.  However,
until recently \cite{Kerr11} it has been impossible to achieve this type of 
quantum turbulent state numerically when the simulation starts from only
two interacting vortices. Which is also true of the classical case.  The
first goal of the calculations reported here was to modify the successful
quantum initial condition of very long anti-parallel vortices \cite{Kerr11} for
the Navier-Stokes equations, then to see if this could also evolve into
a turbulent state.

\subsection{Equations}

The respective classical and quantum equations are as follows. 
First their respective velocity equations (\ref{eq:Euler1},\ref{eq:GP1}) are given,
which show the similarities between the two dynamical systems. Much of the
mathematical analysis of these equations is done with the vorticity
$\bomega=\nabla\times\bu$ and the wave function 
$\psi=\sqrt{\rho}e^{i\phi}$, whose equations are (\ref{eq:Euler2},\ref{eq:GP}) 
respectively, and do not resemble one another. For classical systems either
(\ref{eq:Euler1}) or (\ref{eq:Euler2}) can be integrated in time, for the
quantum case only (\ref{eq:GP}) is integrated.
\vspace{2mm}

\hspace{-5mm}
\bminil{.38}\noindent
Incompressible {\bf Navier-Stokes/Euler}: 
\EQL{eq:Euler1} \hspace{-8mm} \frac{\p\bu}{\p t} + ({\bu}\cdot\nabla){\bu} 
= -\nabla p + \underbrace{\nu\triangle{\bu}}_{\rm dissipation}\EN
\vspace{-1mm}
using $\rho\equiv1$ or $\nabla\cdot{\bu}=0$ \vspace{1mm} \\
What can be integrated: vorticity $\bomega=\nabla\times\bu $ 
\EQL{eq:Euler2}\hspace{-6mm}
\frac{\p\bomega}{\p t} + 
\underbrace{(\bu\cdot\nabla)\bomega}_{\rm advection} =
\underbrace{(\bomega\cdot\nabla)\bu}_{\rm vortex~stretching} 
+ \underbrace{\nu\triangle\bomega}_{\rm dissipation}
\EN
\emini ~~~~~~~~~~
\bminir{.6}\vspace{-4mm}
\begin{center}{\bf Gross-Pitaevskii}\end{center}\vspace{-3mm}
\EQL{eq:GP1} \hspace{-10mm} \rho\left(\frac{\p\bu}{\p t} + ({\bu}\cdot\nabla){\bu}\right)
=-\nabla p + \nabla{\bSigma}\quad{\rm and}\quad 
\frac{\p\rho}{\p t} + \nabla\cdot(\rho{\mathbf u}) = 0 \EN
\vspace{-2mm}
where $\bu=\nabla\phi$, $p=0.25\rho^2$, $\Sigma=$ a quantum tensor.  \vspace{1mm}\\
What is integrated: $\psi=\sqrt{\rho}e^{i\phi}$. 
\vspace{-0mm}
\EQL{eq:GP} \frac{1}{i}\frac{\p\psi}{\p t} = 0.5\nabla^2\psi + 0.5\psi(1-|\psi|^2) \EN
\emini

\medskip
The essential difference between the two systems is that the Navier-Stokes 
equations are dissipative, while the Gross-Pitaevskii equations are an energy 
conserving Hamiltonian system. 

\subsection{Initialisation}

For the three-dimensional incompressible equations, while vorticity is calculated 
by taking derivatives of the velocity, one can still initialise 
with vorticity by inverting its fields to generate the
initial velocities that are usually integrated in time.  
However, this is easier said than done.

\begin{figure}
\label{fig:BustaKerrProfs}
%\includegraphics[scale=.27]{vort11/v07a323232oxz1.jpg} 
%\includegraphics[scale=.23]{v11d/dNSv/v11d128128128nu0p002v2oxzT0f.jpg} 
%\caption{(a) 
\includegraphics[scale=.20]{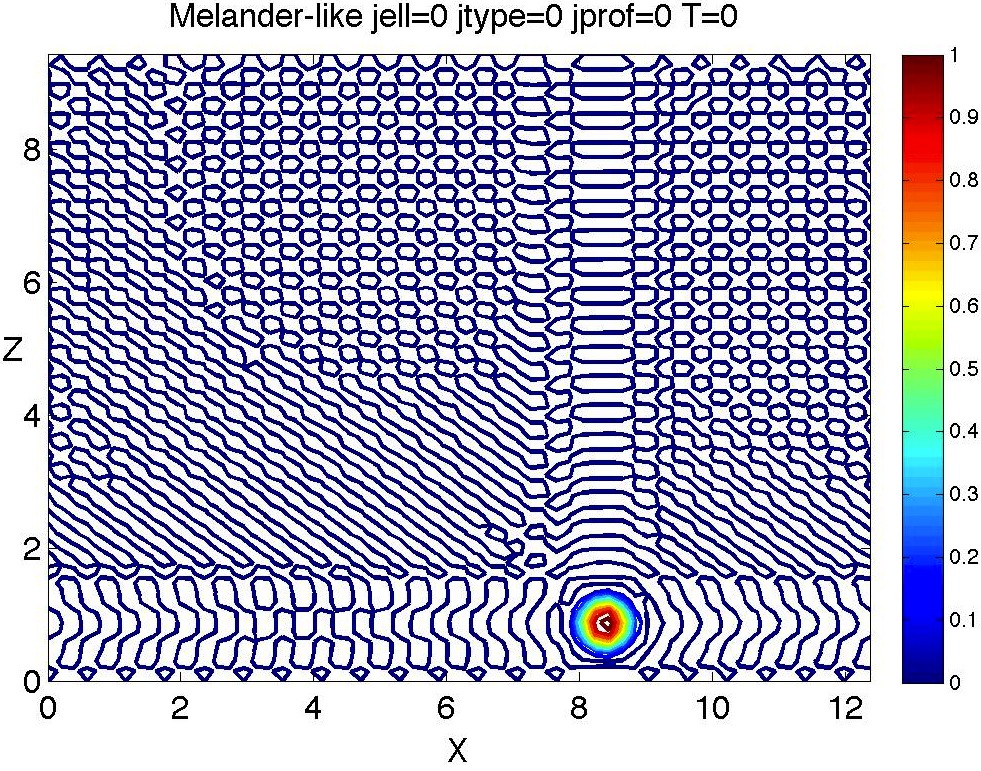}  
\includegraphics[scale=.20]{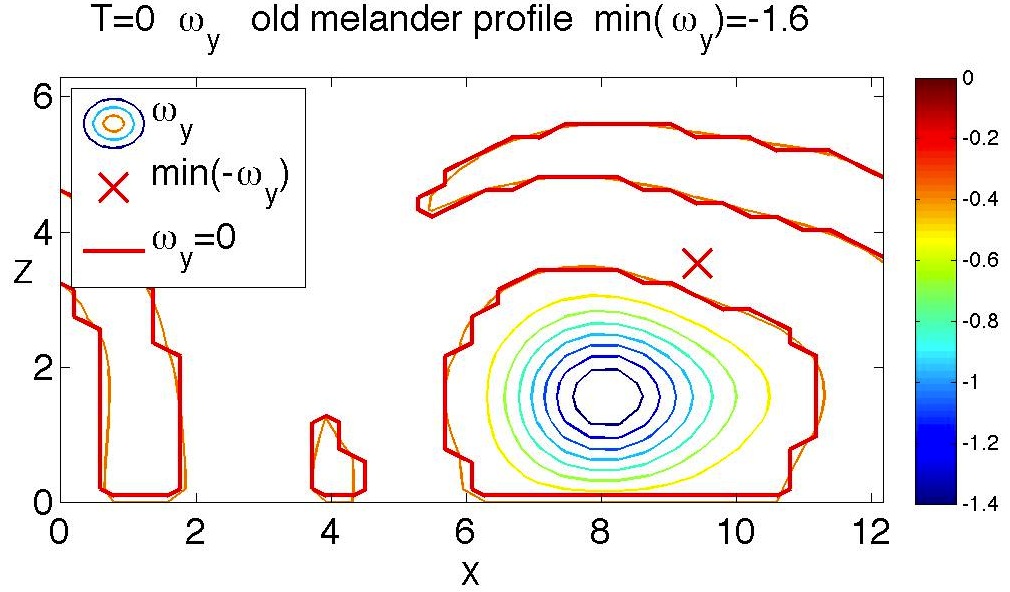}  
\caption{{\bf Left:} Noise generated by one of the original anti-parallel
initial conditions \cite{MelanderH89}. {\bf Right:} 
From \cite{BustaKerr08}. As indicated by the position of the $\min(\omega_y)$.
As in \cite{Kerr93}, $r_\perp$ is used along with a squared-off Gaussian
profile, which is then lengthened  in $x$, in this case by using 
an anisotropic filter in $k_x$.  In \cite{Kerr93} the profile was lengthened 
by adding a localised $\p u_x/\p z$ shear.} 
\end{figure}

These issues underlie some of the recent controversy over whether the Euler equations 
can, or cannot, have singularities, as discussed in the issue of Physica D containing 
the proceedings of the 2007 Euler 350 meeting, including \cite{BustaKerr08}. The
problems with earlier anti-parallel initial conditions, going back to 
\cite{MelanderH89}, are the appearance of negative spots of circulation in
the primary symmetry plane, as shown by figure \ref{fig:BustaKerrProfs}(left).
The consequences of those negative spots discussed in \cite{BustaKerr08} and
the older conferences proceedings it mentions.  Simple fixes such as 
in figure \ref{fig:BustaKerrProfs}(right) have not provided satisfactory corrections.
The inspiration for the new initial condition outlined here only came about by 
attempting to apply an inverted classical initial condition to quantum fluids,
which was then replaced after it developed unphysical instabilities \cite{Kerr11}. 
The three primary elements of the new initial condition are these: 
\vspace{-2mm}
\ITM
\item[i)] A new profile of the vorticity distribution in the core that is
based upon the Rosenhead regularisation of a 2D point vortex: 
\EQL{eq:2DvRosenhead} \bv(\bx)=\Gamma
\frac{(-(y-y_0),x-x_0)}{(x-x_0)^2+(y-y_0)^2+a^2}\quad{\rm and} \quad
\omega=\nabla\times\bv=\Gamma
\frac{a^2}{((x-x_0)^2+(y-y_0)^2+a^2)^2}\,. \EN
Compare this profile with the Fetter Pad\'e approximate for the density 
(\ref{eq:rhoFetter}). Similar except this has been taken to the fourth power. 
This smooth profile behaves
differently than the squared-off Gaussian profile of vorticity introduced by 
\cite{MelanderH89} and used too many times since 
\cite{Kerr93,HouLi06,BustaKerr08}.
\item[ii)] A new direction algorithm whose goal is to find the nearest
position $(x_s,y_s,z_s)$ on the given analytic trajectory to a given 
$(x_i,y_j,z_k)$ on the three-dimensional grid. 
The direction of the vorticity at the points $(x_i,y_j,z_k)$ is then given by
the tangent of the chosen trajectory at $(x_s,y_s,z_s)$.
\item[iii)] Making the vortices very, very long to minimise boundary effects
and allow several reconnection events.  A new goal is to have 
a completely local perturbation that asymptotes into a straight line.  In this way, 
if the vortex line is lengthened, the perturbation remains unchanged.
The double cosine path of the past Euler work \cite{Kerr93,HouLi06,BustaKerr08}
did not satisfy these conditions.  In contrast, the following analytic trajectory does.
\EQL{eq:traj} \bx_\omega(x,y,z) 
=\left(\delta_x\Bigl[2/\cosh\bigl([y/\delta_y]^{p_t}\bigr) -1\Bigr],1,0\right)\,. \EN
The following lengths are used: $\delta_x=-1.6$ and $\delta_y=1.25$. $p_t=1.8$
was chosen in order to localise the perturbation near the $y=0$ symmetry plane.
\item The different physics in the quantum case dictates how to choose
the density and phase around the $\rho=0$ vortex cores.  
The density profile $\rho(r)$ must go from $\rho(0)=0$ to $\rho\rightarrow 1$ as 
$r\rightarrow\infty$.  One example is this low-order Pad\'e solution \citep{Fetter66}:
\EQL{eq:rhoFetter} \sqrt{\rho}(r)=r/\sqrt{r^2+2}\quad{\rm or}\quad
\rho(r)=r^2/(r^2+2)\,.\EN
\ITN % ITM0
\vspace{-2mm}
Points i) and ii) resolve the problem with the negative spots, leading to the
final profile in figure \ref{fig:BustaKerrProfs}.
This has been used for anti-parallel vertical vortices
in a stratified fluid, the anti-parallel unstratified vortices mentioned
here, and anti-parallel Euler vortices. Unphysical initial instabilities due 
to small-scale inbalances have been suppressed, without the
extra massaging or squeezing used before \cite{BustaKerr08,Kerr93}, resulting in
calculations where a cleaner and stronger larger-scale instability forms.
%\index[authors]{Kerr, R.M.}
%\index[authors]{Bustamante, M.D.}
Is this sufficient to lead to a
transition to sustained turbulence from vortex interactions where none
had been seen in earlier work? 

The two simulations are used here say it is.  
Removing the symmetries, the full domains 
are $L_x\times L_y\times L_z=4\pi\times32\pi\times4\pi$ and
$L_x\times L_y\times L_z=4\pi\times32\pi\times8\pi$, with $x,~y,~z$
values going as approximately $[0:12,-50:50,-6:6]$ and $x,~y,~z$
going as $[0:12,-50:50,-12:12]$.
% fNSv1: BOXSIZ(1)=-2.0,BOXSIZ(2)=-8.0, BOXSIZ(3)=-1.0, JSLIPY=1, JSLIPZ=1,
% gNSv1: BOXSIZ(1)=-2.0,BOXSIZ(2)=-8.0, BOXSIZ(3)=-2.0, JSLIPY=1, JSLIPZ=1,
The $L_z=8\pi,~\nu=0.001$ calculation is used to show the full spatial development 
of the structures and the $L_z=4\pi,~\nu=0.0005$ case is used to show that a 
full Kolmogorov $k^{-5/3}$ inertial subrange with very strong $-S_u$ fluctuations 
develops.  The important timescales are the same in the two calculations.

The insets show the upper/left quarter domain near the $y=0$ perturbation plane.
The $t=0$ inset shows that the circular cross-section, taken perpendicular
to the direction of vorticity $\hat{\bomega}$ in its centre, has constant width 
along its entire length and the $t=16$ inset shows that large fluctuations of
vorticity about zero seen previously \cite{BustaKerr08,HouLi06} are now avoided.

All of the initial conditions to be discussed then apply a Fourier
filter of the form
\EQL{eq:initfilter} f_4(\bk)=e^{-\delta_4 (k_x^2+k_y^2+k_z^2)^2}\,, \EN
with $\delta_4=0.01$.  This initialisation was done on
modest meshes, typically $128^3$, before those fields are remeshed onto 
the meshes of true interest.

\begin{figure}
\includegraphics[scale=.24]{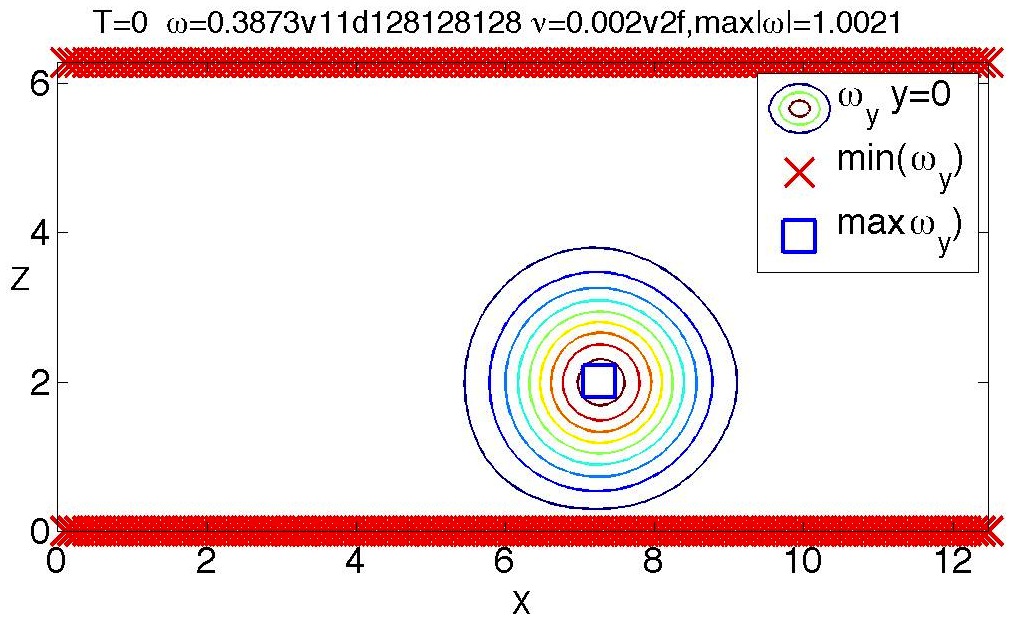} 
\includegraphics[scale=.15,clip=true,trim=0 0 100 100]{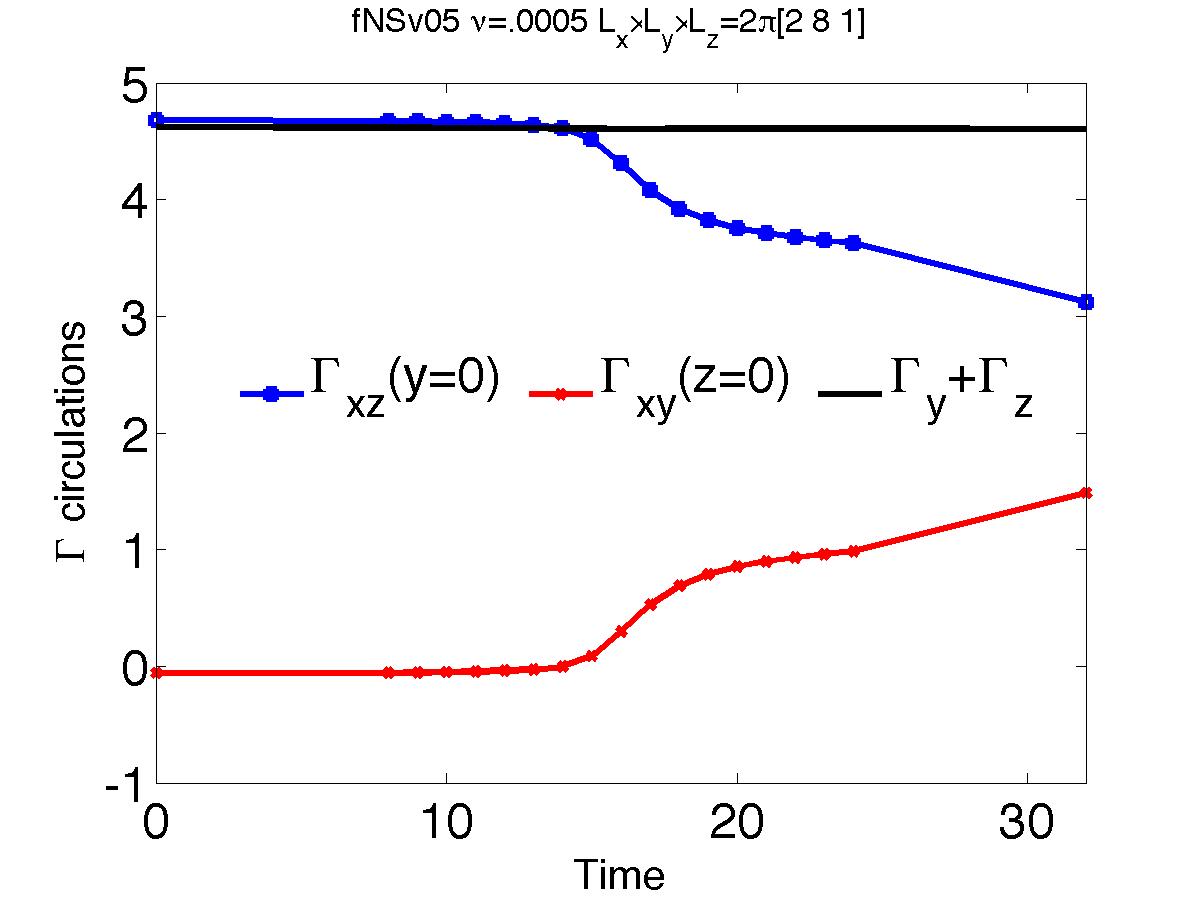}
\caption{{\bf Left:} $x-y$ cross-section of $\omega$ at $t=0$ for $y>0$ 
through the position of maximum vorticity, which is on $y=0$ near $x=7.25$.  
by the continuous crosses of $\omega_y=0$ at these boundaries. 
{\bf Right:} Circulation in the two primary half-symmetry planes.  Their sum remains
constant through the reconnection process which results in nearly half of the original
circulation being reconnected.}
\label{fig:T0oxy11fNSv05circ} 
\end{figure}

\section{Instability steps}

{\bf Initial instability.} The primary inviscid linear instability on anti-parallel 
vortices was originally identified by \cite{Crow72}.  
This quickly becomes nonlinear with the perturbation
in anti-parallel vortices in figure \ref{fig:GPNSreconnect} evolving into
the structures GP $t=1.5$ and NS $t=16$ structures that are beginning to
reconnect at the point of closest approach while generating waves away from those 
points.  The insets at $t=0$ and NS $t=16$ show the perspective
that has appeared in a number of earlier papers, in particular \cite{Kerr93},
with a cut through the symmetry plane.  This cut shows familiar features including
the leading heads, remnants of the original vortices, 
and the following tails, flattened vortex sheets. These sheets are
why the extra resolution is needed in $z$.  

\begin{figure}
\label{fig:T32} 
\includegraphics[scale=.24]{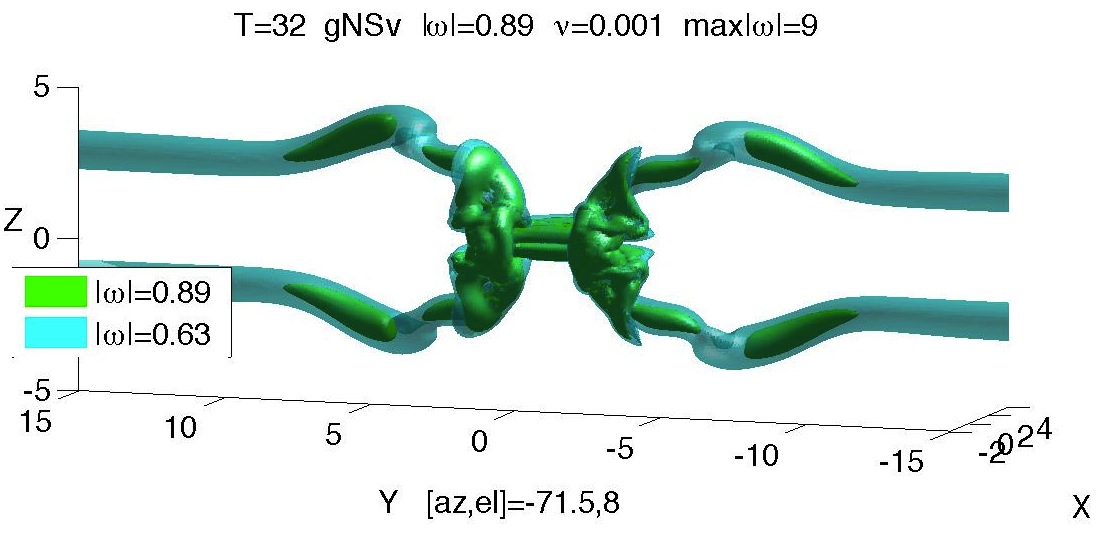}
%\bminil{0.55}
\includegraphics[scale=.17]{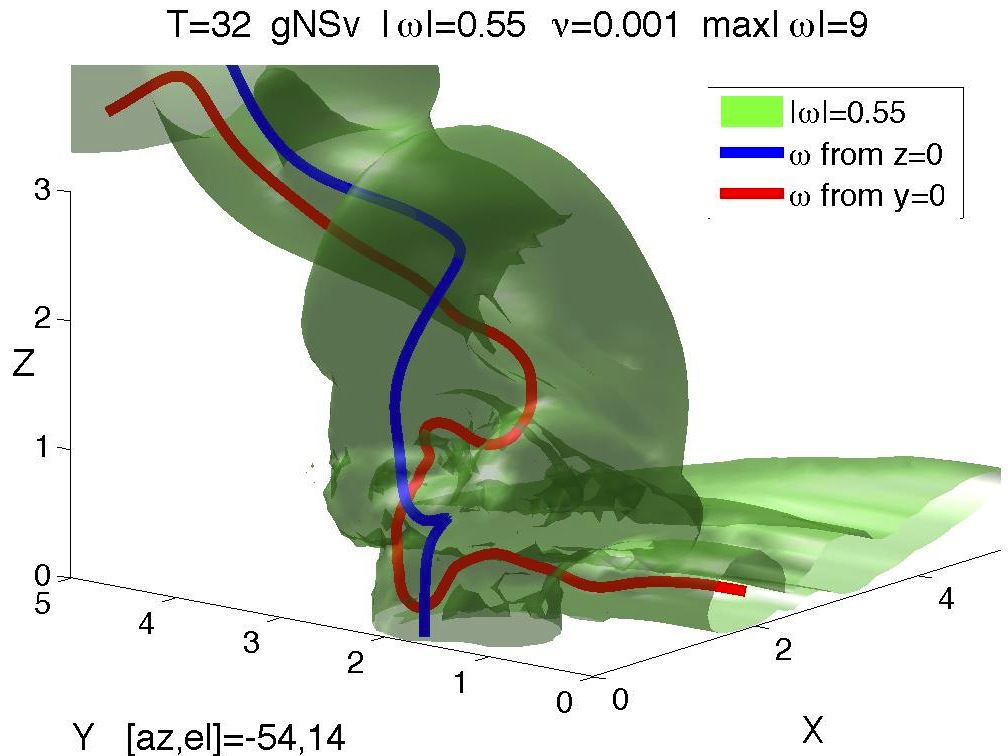} %\emini
%\bminir{0.40}
\caption{$t=32$: Evolved from $t=16$ when reconnection began.  
{\bf Left:} Two isosurfaces. Outer $\omega=.63$
transparent surface in turquoise for comparison with $t=16$ ang later times,
and an inner $\omega=.89$ surface to capture the finer vorticity details.  
{\bf Right:} The $\omega=.55$ isosurface is shown to emphasize how
the tails have started to roll-up, seen on the right of
the cut through the $y=0$ perturbation plane.  How the roll-up connects to
the original vortices is in red while the reconnected parts of the vortices
are blue. They are following each other into the main structure, but
are not yet twisting around one another.}
\end{figure}

After $t=16$, as the reconnection continues, the twists turns into kinks, shown in
figure \ref{fig:T32}.
These twists and kinks do not form if periodic perturbations are used 
\cite{MelanderH89}, which the long domain, plus the localised perturbations, promote.

Ongoing analysis of the curvature for $t<16$ has found a complicated structure of peaks and dips along the vortices with the curvature on the $y=0$ perturbation plane remaining small for all times. These peaks and dips are within the isosurface shown in $t=16$ frame and are not directly connected to the bulges to larger $|z|$ near $|y|=5$. These bulges are more directly related to the requirement that any volume that has been pulled away from the $y=0$ perturbation plane by vortex stretching must go somewhere, and in this case it goes into the bulges.  Another feature of the bulges is how they bend back towards the $z=0$ dividing plane, overshooting the original $z$ positions of the vortices.

By $t=32$, the first reconnection event has completed, with the {\it heads} reconnecting across the $z=0$ plane to form {\it bridges} and kinks that develop from the
inviscid twist generated before $t=16$. 

{\bf Roll-up and twist} 
As the first reconnection is finished at $t=32$, the {\it tails} roll-up to 
form new tubes on the $y=0$ perturbation plane.
These new tubes are commonly known as {\it braids} or {\it threads}, with one indicted 
by the red vortex line coming out of the $y=0$ perturbation plane in the $t=96$ frame. 
To be able to observe the complicated interactions at these two times, two vorticity 
isosurfaces are used: $|\omega|=0.89$ and $|\omega|=0.63$, shown only for the $y>0$ 
half-domain so that the $y=0$, $x-y$ cut is visible.

The two important features highlighted by the $|\omega|=0.89$ isosurface at $t=32$ are 
the roll-up of the {\it tails} and the initial twisting between the {\it bridge} 
and {\it tails}. At $t=32$, the new {\it threads} tuck behind the reconnected 
{\it bridge} at $y\approx2$ and $z=\pm1$ with the two sets of vortices are aligned 
orthogonally, similar to what has been seen after vortex sheets first reconnect in 
simulations initialised with only a few small wavenumber Fourier modes 
\cite{Brachetetal83,HolmKerr07}. % Brachetetal: $t=5$ Figure 16e
Beyond the orthogonal crossing point, the unreconnected {\it threads+tails} begin 
to wrap around the reconnected {\it bridges}, near $y\approx 3$, generating a twisted 
structure that has not been unraveled further.  

The lower $|\omega|=0.63$ surface shows the relationship between these intense regions and the new path for the envelope of the original vortices. This includes the bulge that goes to $|z|\geq 4$ near $y=5$. Note the new intensification of vorticity inside the $|\omega|=0.63$ isosurface for $5\leq |y| \leq 10$. This growth is being driven by a new anti-parallel interaction that is developing around $8\leq |y| \leq 12$. This new source of vortex stretching plays a role in the next stage illustrated
by the frame at $t=96$.

\begin{figure}
\includegraphics[scale=.19]{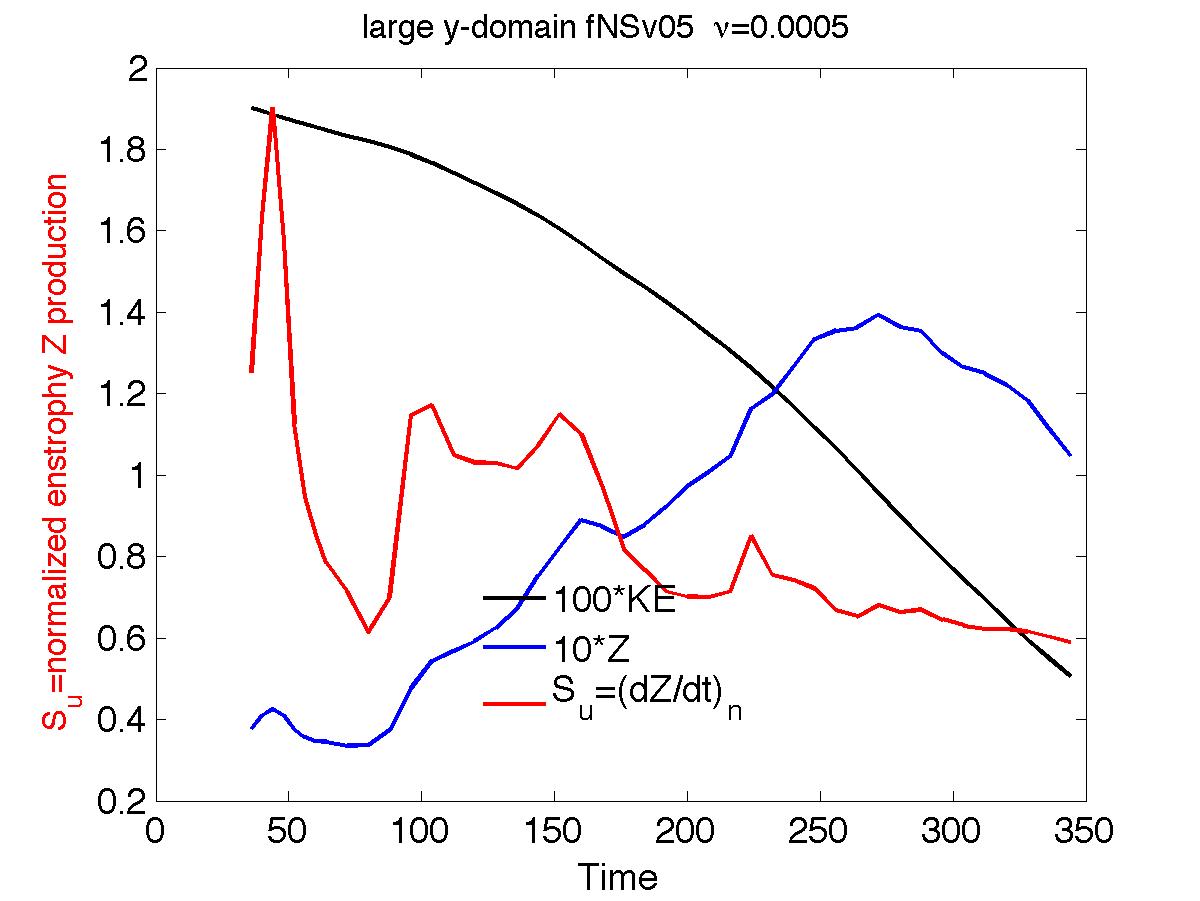}
\includegraphics[scale=.19]{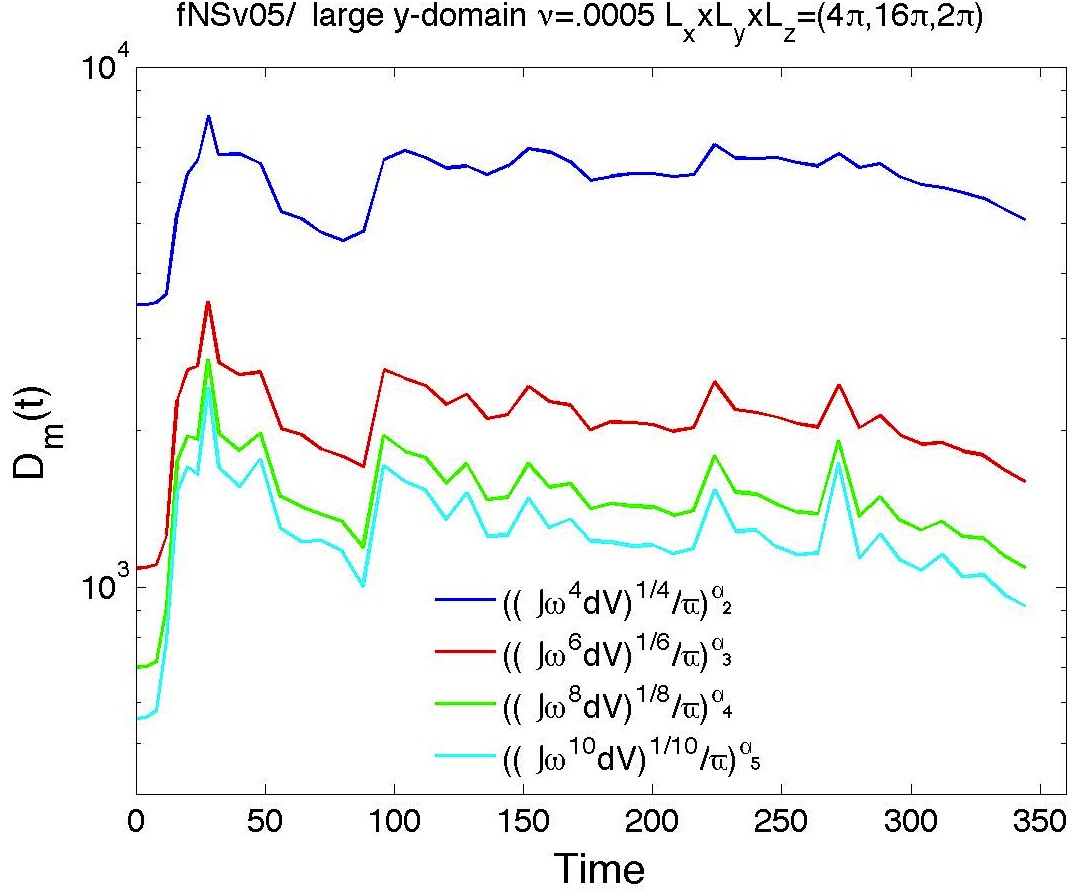} 
\caption{Vorticity and energy time evolution for the $\nu=0.0005$, $L_z=2\pi$
calculation. {\bf Left:} Decay of kinetic energy (lower/black), growth until $t=320$
(then decay) of enstrophy and
velocity derivative skewness: $-S_u$. Several intermediate timescales set using
the growth then decay of the enstrophy $Z$ and $-S_u$, its normalised production.
First timescale: $t_1\approx 48$, showing the largest $-S_u$ and representing 
the completion of the first reconnection, which began about $t_c\approx 14$. 
$t_2\approx 110$ is completion of the second
reconnection and detachment of the first vortex rings.
The $-S_u\approx2$ overshoots the steady-state value of about $-S_u=0.5$ to 0.7
by an unprecedented margin compared to earlier work.  But is not due to 
singular growth, which Euler calculations say should end
after $t\approx 14$.
{\bf Right:} Temporal variation in the higher-order vorticity moments, normalised
and scaled in the manner suggested by \cite{Gibbon11}.  By using this scaling
on this modestly large Reynolds number Navier-Stokes calculation, one can see
that each higher-order moment is bounded from above by the lower-order moments.
}
\label{fig:growth}
\end{figure}

{\bf Timescales} In order to interpret the time evolution shown by the series of frames in figure \ref{fig:GPNSreconnect}, diagnostics are needed for identifying 
timescales. One comes from normalising the production of enstrophy $Z$ 
into the following skewness: 
\EQL{eq:Su} -S_u=c\frac{\overline{\bomega\bS\bomega}}{Z^{3/2}}=
c\frac{dZ/dt}{Z^{3/2}}\quad{\rm where}~
[\bS]_{ij}=\frac{1}{2}\left(\frac{\p u_i}{\p x_j}+
\frac{\p u_j}{\p x_i}\right)~{\rm and}~c~\mbox{is an isotropy coefficient}\,. \EN
It is known from experiments and simulations that at large Reynolds numbers
$-S_u\rightarrow 0.5-0.7$, so the growth and relaxation of $-S_u$ to this
asymptotic value can used to set the timescales of the calculation.

Figure \ref{fig:growth} shows the most significant timescales appearing in 
the global diagnostics.  
On the left is the evolution of the global enstrophy $Z$, its normalized 
production $-S_u$ \eqref{eq:Su} and the decay of kinetic energy. 
On the right is the evolution of the rescaled vorticity moments $D_m$ \eqref{eq:Dm}. 
The $D_m$ are characterised by a hierarchy where the lower-order $D_m$ 
bound the higher-order $D_m$ for all times and order. % 
This hierarchy, including comparisons of these curves to their counterparts
from high-Reynolds number homogeneous, isotropic simulations, is being
investigated further.  
What is important for the evolution described here are the following.
First, the growth of the $Z$, $-S_u$ and the $D_m$ are nearly identical up 
until $t\approx16$. This is when the first reconnection begins 
and is a candidate time for when there could be a singularity 
in the $\nu=0$ Euler limit.

The next important time is $t\approx48$ when $-S_u$ has its first and largest peak for both cases and there is a minor peak in $Z$. Isosurfaces at $t=48$ are almost the same as those at $t = 32$, with a little additional twisting up the reconnected {\it bridges}. $-S_u$ decreases rapidly for $t>48$, reaching a minimum at $t\approx 90$, which is when the second reconnection is beginning and is discussed next.  The additional peaks in $-S_u$ can all be associated with the additional reconnection events being discussed.  These features are largely mirrored by the $D_m$ in the right frame of figure \ref{fig:growth}, including $D_\infty$.

{\bf Spirals} By $t=32$, the reconnected {\it bridges} have begun 
to separate in $y$. This is because the curvature underlying the Biot-Savart 
interaction reverses its direction in the reconnected vortices,
so the interaction goes from attraction 
in $z$ for $t<16$ before reconnection, to repulsion in $y$ for $t>32$ after 
reconnection.  As the $t>32$ vortices separate, the reconnected {\it bridges} wind the 
unreconnected {\it threads} about themselves, illustrated at $t=96$ by how the 
red {\it thread} coming from the $y=0$ plane snakes around the blue line coming from 
the $z=0$ {\it bridge}. For large $y$, the red and blue lines meet and joint into the 
original vortex. 

The other major feature to note in the $t=96$ frame is that the new anti-parallel attraction seen for $8\leq y\leq 12$ at earlier times has now moved out to $y\approx 17$ and has progressed to the stage where the vortices are starting to touch and reconnect again.  The inner isosurface shows flattening, curvature, and preliminary roll-up similar to what forms at $y=0$ at $t=16$, just before the first reconnection. Furthermore, for the period just before and after $t=96$, the normalised enstrophy production and the higher-order $D_m$ in figure \ref{fig:growth} all grow significantly, as they do around $t=16$, the time of the first reconnection.

After the reconnection that starts at $t=96$ has been completed, we are left with a disconnected vortex ring with strong spiraling patterns snaking completely around it, as shown at $t=176$.  By $t=176$, a third reconnection site is developing near $y\approx35$, which leads to a second vortex ring disconnecting, with new spirals forming on its low $y$ side.  A second ring with spirals then disconnects, as shown at $t=256$.  Given a large enough domain in $y$, this process would probably continue to produce new spiral rings, as in the quantum vortex case \cite{Kerr11}. It also has similarities to experimental observations such as \cite{LewekeWilliamson11} where thin {\it threads} connecting a chain of vortex rings are seen.

\section{Spectra and moments}
\begin{figure}
%\bminil{0.55}
%\includegraphics[scale=.32]{v05KEkyT44-344.jpg}
\includegraphics[scale=.32]{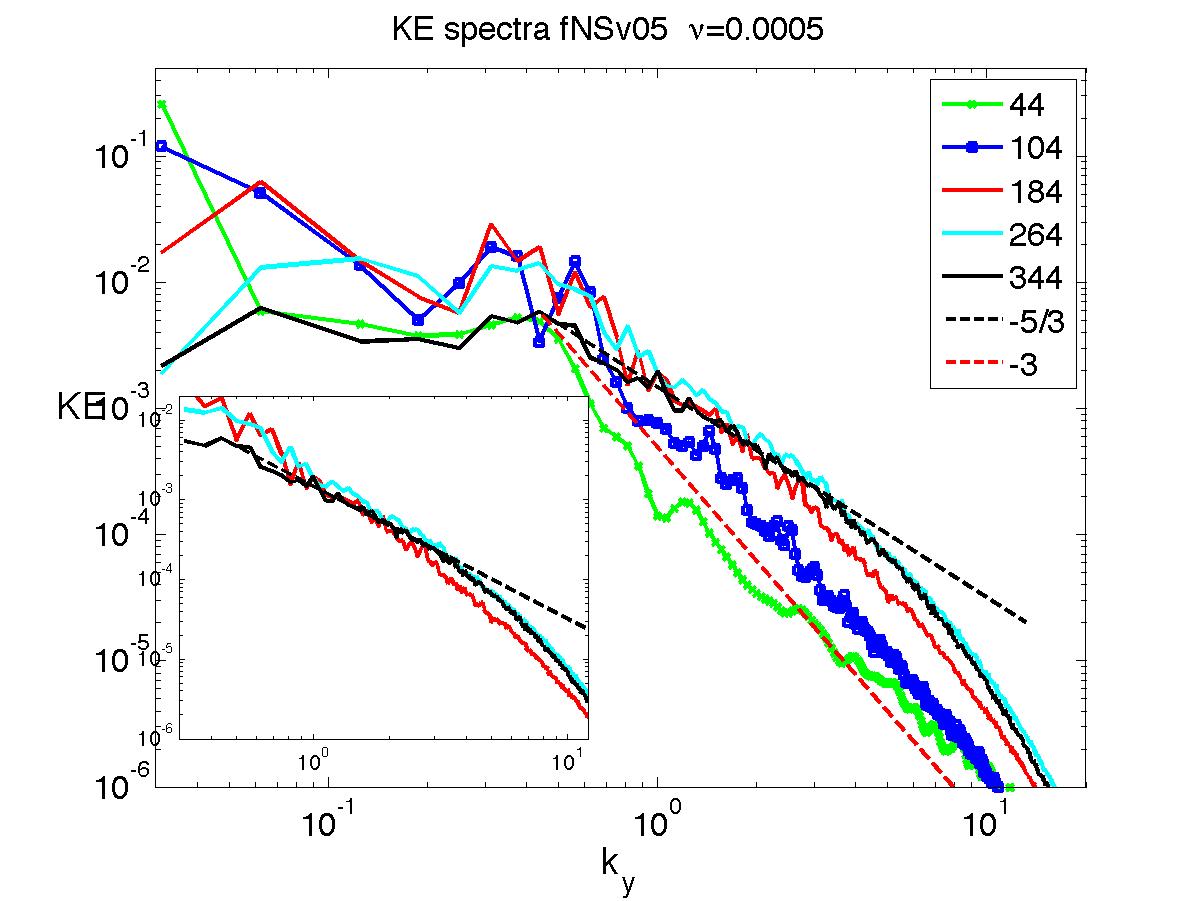}
%\includegraphics[scale=.14]{v11d/fNSv05/T352-384-416KEkynormline.jpg}
%\emini\bminir{0.42}
\caption{Kinetic energy spectra: Domain: $2\pi(2,8,1)$ $\nu=0.0005$. 
%Full, normed and scaled in $k_y$, $k_x$ and $k_z$.
Uncompensated spectra shows the progression from
a $k^{-3}$ spectrum at short times to the formation of a $k^{-5/3}$ at later times.
%Right: Compensated similarity spectra.
As the large-scale, low wavenumber part of the spectrum begins to decay, Kolmogorov 
similarity (using the time dependent dissipation $\epsilon(t)=\nu Z$) 
sees convergence starting at $t=184$.
The estimated Kolmogorov coefficient $K_0$ is between
$K_0$=1.05 and 1.1.  This is lower than what is obtained from large forced 
numerical simulations, which typically give $K_0\approx 1.5$, perhaps because 
not all of our global enstrophy $Z$ (and therefore dissipation 
$\epsilon=\nu Z$) is participating in the turbulent dynamics.  
} %\emini
\label{fig:spectrafNSv05}
\end{figure}

The initial high-wavenumber spectra are given by the hyperviscous filter,
which quickly develop into power laws that are steeper than $k^{-4}$,
as seen in Euler calculations \citep{Kerr93}. The subsequent spectral evolution 
is shown in figure \ref{fig:spectrafNSv05}, starting at $t=44$.
Eventually, starting at $t=184$, a $k^{-5/3}$ begins to form, lengthen, 
and collapse according to Kolmogorov similarity.  $k_x$ and $k_z$ spectra have
also been determined. They do not have the same 
extent as the $k_y$ direction because the physical space domains in those
directions are smaller, but their inertial subranges
are consistent in magnitude and extent with those in the $k_y$ direction.  

While spectra can provide some measure of whether a fluid is turbulent or not,
they provide little information about the statistical nature of physical space
velocity fluctuations and intermittency.  % These limitations on the value
%of spectra are even more true for experiments, where single or double variable 
%spectra in one spatial direction are usually all that are feasible.  
If the experimental conditions are relatively steady for long periods 
of time, there are a number of alternative physical space statistics that 
can be used to characterise the intermittency of turbulent flows.  However,
determining many of these numerically is restricted
by how rapidly they converge to steady statistics.
In the analysis of high Reynolds number, very large numerical simulation, what
\cite{Yeungetal12} have demonstrated is their ratios converge, even if their
values don't. As recently pointed out \cite{Kerr12}, this can be explained
by the mathematics in \cite{Gibbon11}.

What insight can these reconnection calculations provide? They can if the rise and fall of these intermittent statistics can be tied to the reconnection events and the growth of enstrophy production.

First, is intermittency growing sufficiently rapidly as the Reynolds number increases for these reconnection simulations to be useful? 
The strong growth in the skewness fluctuations in figure
\ref{fig:growth}(left) suggests that this is the case as the two initial peaks near $t=40$ and $t=105$ double when the Reynolds number is doubled. 
Even higher growth is expected when this is redone with an $L_z=4\pi$ domain.

Which brings us to the rescaling of the higher-order vorticity moments as
a function of time in the manner recently suggested by \cite{Gibbon11}, 
given in figure \ref{fig:growth}(right). This figure shows a hierarchy of 
higher-order vorticity moments, higher than $m=1$ (the enstrophy), 
which would off the upper scale at the top.  The normalisation is based upon 
the frequency scale $\varpi=\nu/L_z^2$ and a power law designed to compare 
the degree of intermittency for each moment and goes as
\EQL{eq:Dm} D_m=(\varpi H_m)^{1/\tilde{\alpha}_m}\quad{\rm where}\quad
\tilde{\alpha}_m=4m/(4m-3) \,,\EN
where the $H_m=(\int\omega^{2m}dV)^{1/2m}$ are the standard vorticity moments.
With this scaling, each of the higher-order moments is bounded from above by the 
lower-order moments. Furthermore, as the order increases, the moments become
bunched in regions of strong growth.  The ordering is fostered by the choice of
frequency scale and the bunching is fostered by the choice of power law.  The
maximum of vorticity, the $m=\infty$ norm, follows the same scaling and appears
just below the $m=5$ curve.

There is no rigorous proof that this ordering should be obeyed, but this observation
should be significant for the question of 
whether the Navier-Stokes equations have singularities or not.
First, could it be shown that for any sufficiently smooth Navier-Stokes initial condition that these moments relax quickly into this ordering, and that this ordering is then maintained? If so, then we would know that all the higher-order vorticity norms will eventually be bounded from above by the lower-oder vorticity norms. Which brings us to the lowest order, $m=1$ norm, the enstrophy. Implying that if the growth of enstrophy could be bounded for all times, then all higher-order moments of vorticity could be bounded, 
including $H_\infty$. 

\section{Summary}

The case is made in this paper that if the initial vortices are well-balanced and the domain is large enough for two independent reconnection steps to form, then the flow wil rapidly break down into three-dimensional, fully- developed turbulence.

The initial configuration used to demonstrate this consisted of very long, anti-parallel vortices. It is argued that the long domain is needed in order to accommodate the multiple reconnections, which enhance vortex stretching rates and the generation of small-scale vortex structures within the vortex rings. In addition to making the initial vortices very long, several new features are applied during initialisation that are designed to make it less likely that the flow will generate vortex sheets. These include a new balanced profile and an improved way of mapping the direction of the vorticity onto the three-dimensional mesh.

To get to the turbulent state, the vortices progress through the following steps: First, until the first reconnection, the vortex dynamics are largely consistent with existing work on possible singularities of the Euler equations. This includes a head and tail structure with the maximum of vorticity between these them. A feature not clearly noted before is how the vortices twist back upon themselves towards secondary points along the original anti-parallel vortices. This nonlocal amplification of the anti-parallel interaction provides extra stretching along the entire vortices, including the primary symmetry plane.

Jumping to the last figure of higher-order moments, note the convergent growth of these moments leading up to $t=16$. This is related to singular Euler dynamics that is being discussed elsewhere.

The next step is reconnection on the primary symmetry plane. Only the half associated with the head reconnects into two ``bridges'', leaving the tails to roll-up into two ``threads''. Together these form a locally orthogonal configuration with steep energy spectra and minimal energy dissipation.

The third step starts with orthogonal configurations and is modified, compared to past simulations, by the stretching induced by the anti-parallel interactions at the secondary points. Due to this secondary stretching, as the threads begin to wrap around the newly reconnected bridges, they are also pulled along the bridges to form spirals.

Eventually there are two more reconnections at secondary points, which result in vortex rings separating from the original vortex lines. It is only at this stage, while the spectra are becoming less steep, that significant dissipation develops, first at higher wavenumbers.

The fourth step will be defined as an extended period with multiple reconnections, forming multiple vortex rings, during which a nascent -5/3 extends to low wavenumbers and additional statistics achieve values consistent with fully-developed turbulence. Each reconnection is at the secondary interaction positions induced by the previous anti-parallel interaction and reconnection. And each pair of reconnections forms a vortex ring.

In the calculations reported here, there are three distinct reconnection events from which two sets of independent vortex rings form on either side of the symmetry plane. 
For the quantum vortex calculations in figure \ref{fig:GPNSreconnect},
it is seen that there is no end to the number of rings that can form, so long as the domain is large enough. For each set of reconnections, additional stretched spirals form as the $k^{−5/3}$ energy spectrum appears. The final vortex structure is a chain of vortex rings connected by thin threads, as seen in some experiments.

The final state satisfies all the usual properties of fully-developed turbulence, usually associated with homogeneous and isotropic flows, neither of which is true for these simulations. There is a Kolmogorov $k^{−5/3}$ subrange with roughly the correct constant. And after highly intermittent variations, the velocity derivative skewness $-S_u$, 
or normalised enstrophy production settles to values of about 0.5 to 0.6, which are the values characteristic of fully-developed turbulence. It is then shown that the intermittency of $-S_u$ can be extended to intermittency in higher-order moments of vorticity. Moments that we hope will be determined numerically in some upcoming cutting-edge simulations of forced, homogeneous, isotropic turbulence similar to that used by \cite{Yeungetal12}.

\section*{Acknowledgements}

All calculations were done on clusters at the University of Warwick’s Centre for Scientific Computing. Travel support from COST-Aerosols and Particles is appreciated. I thank J.D. Gibbon for communicating his latest mathematics results on vorticity moments.

\section*{Acknowledgements}
The author thanks the organisers of this meeting for the opportunity to speak.
Computing support was provided by the Warwick Centre for Scientific Computing.

\end{document}